\documentclass[preprint]{aastex}

\newcommand{\brs}{{\it brightenings}}
\newcommand{\sdss}{SDSS\,1238}

\shorttitle{SDSS J123813.73-033933.0, a CV evolved beyond period minimum}
\shortauthors{A. Aviles, S. Zharikov, G. Tovmassian et al.}

\begin{document}


\title{SDSS J123813.73-033933.0, a cataclysmic variable evolved beyond the period minimum.
}

\author{A.~Aviles\altaffilmark{1}, S.~Zharikov\altaffilmark{1}, G.~Tovmassian\altaffilmark{1}, R.~Michel\altaffilmark{1}, M. Tapia\altaffilmark{1},}
\affil{Instituto de Aatronom\'{i}a,
Universidad Nacional Aut\'{o}noma de M\'{e}xico,
Apartado Postal 877, 22800, Ensenada, BC, M\'{e}xico}

\author{M. Roth\altaffilmark{2},}

\affil{Las Campanas Observatory, Carnegie Instututio of Washington,
Casilla 601, La Serena, Chile}
\author{V.~Neustroev\altaffilmark{3},}

\affil{Centre for Astronomy, National University of Ireland, Galway, Newcastle Rd., Galway, Ireland}

\author{ C.~Zurita\altaffilmark{4},} 

\affil{Instituto de Astrofisica de Canarias, c/ via Lactea s/n, La Laguna, E38200, Tenerife, Spain}

\author{M.~Andreev\altaffilmark{5}, A.~Sergeev\altaffilmark{5}, }

\affil{Institute of Astronomy, Russian Academy of Sciences, Terskol, Russia}

\author{ E.~Pavlenko\altaffilmark{6},}
\affil{Crimean Astrophysical Observatory, Nauchny, Ukraine}

\author{V.~Tsymbal\altaffilmark{7},}
\affil{Tavrian National University, Dep. Astronomy, Simferopol, Ukraine}

\author{ G.C.~Anupama\altaffilmark{8}, U.S.~Kamath\altaffilmark{8},  D.K.~Sahu\altaffilmark{8}}

\affil{Indian Institute of Astrophysics -- CREST, Bangalore 560 034, India}


\altaffiltext{1}{Instituto de Aatronom\'{i}a,
Universidad Nacional Aut\'{o}noma de M\'{e}xico,
Apartado Postal 877, 22800, Ensenada, BC, M\'{e}xico }

\altaffiltext{2}{Las Campanas Observatory, Carnegie Instututio of Washington,
Casilla 601, La Serena, Chile}
\altaffiltext{3}{Centre for Astronomy, National University of Ireland, Galway, Newcastle Rd., Galway, Ireland}

\altaffiltext{4}{Instituto de Astrofisica de Canarias, c/ via Lactea s/n, La Laguna, E38200, Tenerife, Spain}

\altaffiltext{5}{Institute of Astronomy, Russian Academy of Sciences, Terskol, Russia}

\altaffiltext{6}{Crimean Astrophysical Observatory, Nauchny, Ukraine}

\altaffiltext{7}{Tavrian National University, Dep. Astronomy, Simferopol, Ukraine}
\altaffiltext{8}{Indian Institute of Astrophysics, II Block Koramangala, Bangalore 560034, India}

\begin{abstract}
We present infrared JHK photometry of the cataclysmic variable SDSS\,J123813.73-033933.0 and analyze it along with optical spectroscopy, 
demonstrating that the binary system is most probably comprised of a massive white dwarf 
with $T_{\rm {eff}}=12000\pm1000$\,K and a brown dwarf of  spectral type  L4. 
 The inferred system parameters
suggest that this system may have evolved beyond the orbital period minimum and is a bounce-back system.  

 SDSS\,J123813.73-033933.0 stands out among CVs by exhibiting 
the cyclical variability  that \citet{2006A&A...449..645Z} called \brs. These are not related to specific orbital phases of the binary system and are fainter than dwarf novae outbursts, that usually occur on longer timescales. This phenomenon has not been observed extensively and, thus, is poorly understood.
The new time-resolved, multi-longitude photometric observations  of SDSS\,J123813.73-033933.0 
allowed us to observe two consecutive \brs\  and  to determine their recurrence time. The period analysis  of all observed \brs\  during 2007 
suggests a typical timescale that is close to a period of $\sim9.3$
hours. However,   the \brs\  modulation is not strictly periodic, possibly maintaining coherence only   on timescales   of several weeks. The characteristic  variability with double orbital frequency that  clearly shows up during  \brs\  is also analyzed.  
 
The Doppler mapping of the system shows the  permanent presence of a spiral arm pattern in the accretion disk. A simple model is presented to demonstrate that spiral arms in the velocity map appear at the location and  phase corresponding to the  2:1 resonance radius and constitute themselves as a double-humped light curves.   The  long-term and short-term variability of this CV is discussed together with the spiral arm structure of  an accretion disk  in the context of  observational effects taking place in bounce-back systems. 
\end{abstract}





\section{Introduction}
\label{sec:intro}

The object catalogued as SDSS J123813.73-033933.0 (hereafter \sdss) was identified with a faint ($r = 17.82$ mag) short-period  cataclysmic variable  (CV) by \citet{2003AJ....126.1499S}. The optical spectrum of \sdss\  shows a blue continuum with broad absorption features originating in the photosphere of a white dwarf surrounding   double-peaked Balmer emission lines,  formed in a high inclination accretion disk.   
The  orbital period of the system is  $P_{orb} = 0.05592(35) d = 1.34(1)h$, based on  spectroscopic data \citep{2006A&A...449..645Z}.
The orbital period and the spectral features match those of  WZ-Sge-type systems, but with this,  similarities practically end. A number of observed aspects of the system differ from the majority of short period CVs.
 The most intriguing  characteristic,  which  we found in this system, is a sudden and  fast rise in brightness  up to  $\sim 0.45$ mag during a short time, of about half of the orbital period.   After reaching its peak, the  brightness slowly decreases,  lasting  $\sim 3-4$ hours, down to the  quiescence level. We call  these events \brs\ in order to distinguish them from the more common outbursts, humps,  flickering and other types of variability documented in short period CVs.  
 These \brs\  seemed to happen cyclically about every  8-12\,hours.  In addition to \brs, a nearly permanent sinusoidal variability was detected  in  the light curve of \sdss\  {with a period  half that of the  spectroscopic orbital   period  $P=P_{orb}/2=40.25min$ (hereafter, the double-humped light curve).}  
The amplitude of the double-hump variability depends  on the phase of the  \brs. It  increases with a total rise in  brightness of  up to $\sim0.2$ mag and decreases  until almost disappears  during the quiescence \citep{2006A&A...449..645Z}.
 A similar behavior was found later by \citet{2006AJ....131..973S} in another short period CV, SDSS\,J080434.20+510349.2 (hereafter SDSS\,0804) which  has an identical spectral appearance to \sdss\  in  quiescence.
\citet{2006A&A...449..645Z, 2008A&A...486..505Z} advanced  the hypothesis that the double-humped light curve is a signature of 2:1 resonance in the accretion disks of these systems. { In order for the accretion disk  to reach permanently  the 2:1 resonance radius,} the mass ratio of the binary component must be extreme ($q \leq 0.1$), and as such these objects could  qualify as bounce-back systems, e.g.  CVs, which are old enough to reach the period minimum and  leap toward slightly longer  orbital periods, as predicted by \citet{1981AcA....31....1P}.  It is supposed that accretion disks of WZ Sge systems reach 2:1 resonance radius during super-outburst, when some of them  have been noted to show double-humped light curves \citep{2002PASP..114.1364P}. The super-outbursts of WZ Sge type systems are infrequent and happen every two dozen or more years.
 Oddly enough, SDSS\,0804 went   into  the super-outburst in 2006 \citep{2007ASPC..372..511P} and exhibited all  necessary attributes of a classical WZ Sge-type object. Regrettably, the \brs\ disappeared from the light curves of SDSS\,0804 after the super-outburst, although the double-hump light curve persists  \citep{2008A&A...486..505Z}.  Thus,   the \sdss\ remains the only object that still shows  \brs.  
 Intrigued by the new photometric phenomenon observed  in these two systems, we conducted a new time-resolved photometric study of \sdss\  to establish the reasons behind their common nature, understand the origin of the cyclic \brs\  and its relation to the amplitude of the double-humped light curve. Meanwhile, we discovered that \sdss\ was marginally detected as an infrared source by 2MASS survey, and secured accurate near-IR photometry of the object.
 In Sect.2 we describe our observations and data reductions. The data analyses and the results are presented in Sects.3,4,5, while a general discussion and conclusions are given in Sect.6. 

\section{Observations and data reduction}
\label{sec:obs}

The object  is listed  in the 2MASS\footnote{http://www.ipac.caltech.edu/2mass/} Point Source Catalogue  with   $J=16.65(13)$, $H=16.49(23)$, $K\sim16.42$
magnitudes. These  magnitudes are  close to  the detection limits of   2MASS, particularly in the $K$-band.  In order 
have more accurate photometry, 
we obtained new 
observations of SDSS\,J123813.73-033933.0   in $J\,H\,K_s$ on 17 June 2009 with the
near-infrared camera PANIC (Martini et al. 2004) attached to the 6.5 m Baade/Magellan
Telescope at Las Campanas Observatory (LCO). PANIC provides an
image scale of 0.125$''$ pixel$^{-1}$
on a Hawaii HgCdTe $1024 \times 1024$ array detector. The FWHM of the point-spread
function was between $0.55''$ and $0.60''$ during our observations. For each filter, 9 dithered
frames spaced  by $10''$  were taken, with total on-source integration times of 540~s
in each of the three filters.  The nine frames were shifted and averaged to produce the
final images. Standard sky-subtraction and flat-field correction
procedures were applied.  Aperture ($1.2''$) photometry was performed with  DAOPHOT
within IRAF in the standard way. Flux calibration was performed using standard stars SJ~9146 and SJ~9157 from the 
list  of \citet{1998AJ....116.2475P} and the total errors are estimated to be less than 0.05 magnitudes. The resulting magnitudes 
of \sdss\ are  $J=17.07(5)$, $H=16.65(5)$, $K=16.42(5)$ and corresponding colors $J-H=0.42$, $H-K_s=0.23$ and $J-K_s=0.65$. 


 In order to investigate whether flux variations in the near-IR  occur in this system in timescales of a few minutes, we measured the
$J$, $H$ and $K_s$ fluxes from each of the nine short-exposure (60~s) frames in each filter. In the  $\sim15$\,min that lasted
each series, we did not detect any variability  within the $\sim 0.15$ mag photometric uncertainty associated with each
single frame. Note, that in longer timescales (comparable to the orbital period), the system  is expected to show some variability in the 
near-IR, mostly due to the elliptical shape of Roche-lobe filling secondary, but these  could be missed in a 15 min time series.
The ellipsoidal variability of the secondary can be calculated and it has been taken into account in further considerations of IR magnitudes.

The objective of our optical photometry  of \sdss\ was to study of phenomenon of  \brs. 
Taking into account the long duration of the \brs\  and the uncertainty of the  cycle period \citep{2006A&A...449..645Z}, we planned and executed a multi-longitude observational campaign of this object. Time-resolved CCD  photometry 
was obtained at several facilities: the 1.5 m  telescope at the Observatorio Astron\'{o}mico Nacional at San Pedro M\'{a}rtir in Mexico;  the 0.8 m IAC80 telescope at the Observatorio del Teide in the Canary Islands, Spain; the 2.1 m telescope at the Bohyunsan Optical Astronomy Observatory (BOAO) in South Korea; the 2m telescope at the Terskol Astrophysical Observatory in the Northern Caucasus, Russia, the 2m Himalayan Chandra Telescope of the Indian Astronomical Observatory (IAO), Hanle, India.
The data reduction was performed using both ESO-MIDAS and IRAF software. The images were bias-corrected and flat-fielded before aperture photometry was carried out. The log of photometric observations is presented in Table.1.

The long-slit observations have been obtained with the Boller \& Chivens spectrograph\footnote{www.astrossp.unam.mx} on the 2.1-m telescope at the SPM site 
with a resolution of 3.03 \AA \ pixel$^{-1}$. The spectra  span the wavelength range 4000-7100 \AA.\   In order to improve the signal-to-noise ratio, we obtained a series of phase-locked spectra: 10 spectra were taken at equal phase intervals over a single orbital period $P_{orb} =80.5$\,min with an exposure time of 486 sec per spectrum. This sequence of spectra was repeated at exactly the same phase intervals for subsequent periods and subsequent nights. This allows us to calculate the phase-averaged spectra, summarizing the spectra of the same orbital phase obtained during one night and the whole set of observations without further decreasing the time resolution.
The log of spectroscopic observations is presented in Table.2.

\section{Spectral energy distribution: system parameters and distance to the object}

\begin{figure}
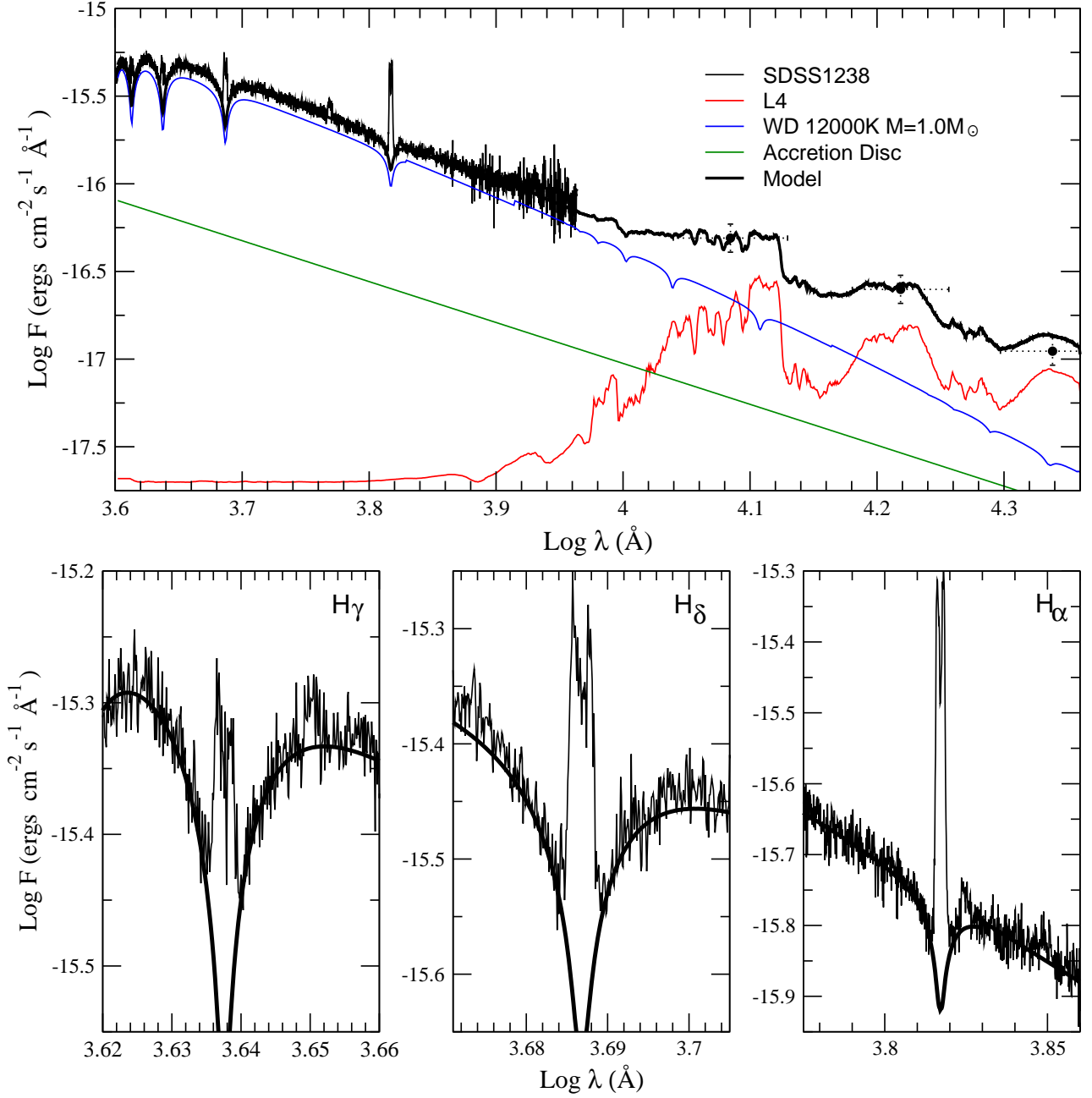

  \includegraphics[width=17cm,bb = 22 47 708 403, clip=]{Avilf01a.eps}
   \includegraphics[width=17cm,bb = 27 50 745 403, clip=]{Avilf01b.eps}
   \caption{The spectral energy distribution of SDSS 1238 and the result of the model fit (top). Spectrum fragments around 
   H$_\gamma$, H$_\delta$ and H$_\alpha$ lines are shown in  bottom panels.} 
   \label{fig:1}
\end{figure}

\begin{figure}
  \includegraphics[width=17cm,clip=]{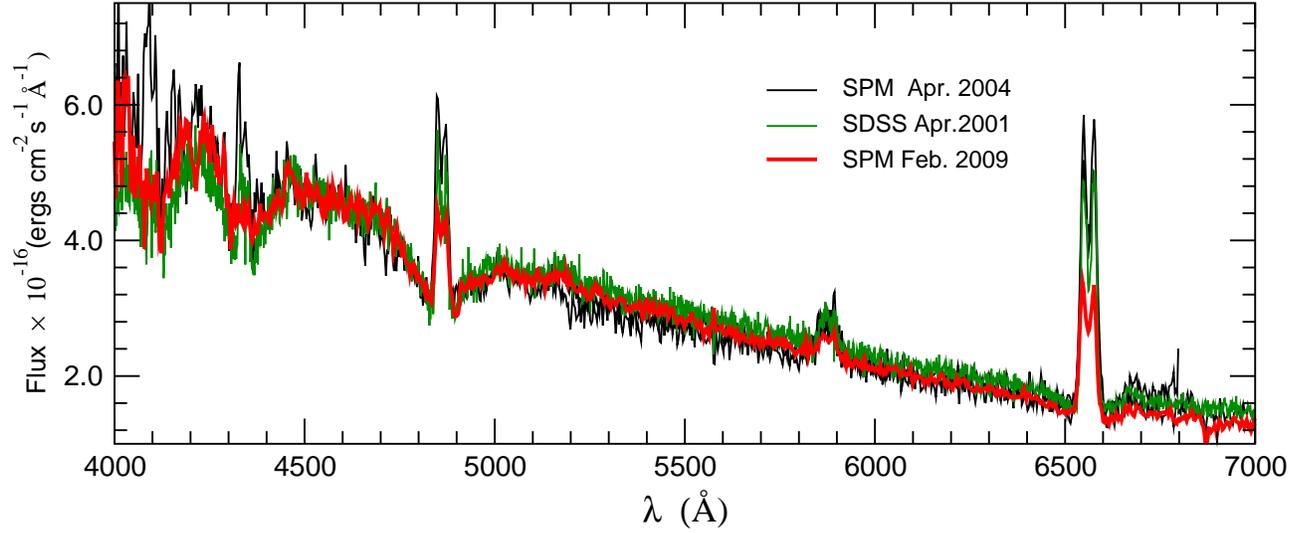}
   \caption{The low-resolution time-average spectra of SDSS1238 obtained in different epochs.} 
   \label{fig:2}
\end{figure}

The spectral energy distribution (SED) of the object  in the range of 4000-25000\AA \ is shown in Fig.\ref{fig:1}. The detailed description of the \sdss\ optical spectrum was given in our previous paper \citep{2006A&A...449..645Z}. The overall appearance of the spectrum has not changed, but we detect a significant variability of  the equivalent widths of  Balmer emission lines from epoch to epoch.
The Balmer lines  are about two times weaker in the 2009 observations compared to the 2004 spectra (Fig.\ref{fig:2}).  The continuum, however, has not  changed during the last five years, as we compare V-band magnitudes  in  quiescence  between \brs.  The average quiescence magnitude\footnote{ Secondary photometric  standards were established in the field of \sdss, by calibrating them using reference star S2003313360 from GSC-II www.gsss.stsci.edu/Catalogs/GCS/GSC2/GSC2.html} between \brs\  remains  constant at around  $V \cong17.8\pm0.1$.

The present near-IR  measurements demonstrated that  there is significant IR excess emission to that expected from  the Rayleigh--Jeans tail of the optical spectrum for a white dwarf (WD). In fact,  neither a WD nor a power-law flux from the accretion disk, nor their combination, can explain  the observed IR excess.   The most probable source of IR  excess is the radiation from the secondary star. 
 To determine the spectral type of the secondary and the distance to the system we fitted the observed optical-infrared  spectral energy distribution  of the object with a simple model: the total  flux  $F^{*}(\lambda)$ is the sum of contributions from a  WD with a  hydrogen atmosphere,   $F_{\rm {WD}}(T_{\rm {eff}}, \lambda)$ (DA type WDs),   
 an accretion disk with $F_{\rm {AD}} \sim \lambda^{-\frac{7}{3}}$ \citep{1969Natur.223..690L}, and  a red/brown dwarf with $F_{\rm {BD}}(\lambda)$: 
\begin{equation}
 F^{*}(\lambda) = F_{\rm {WD}}(T_{\rm {eff}}, \lambda) + F_{\rm {AD}}(\lambda) +  F^{\rm {SpT}}_{\rm {BD}}(\lambda) 
\end{equation}
Brown dwarf fluxes were  taken from the literature \citep{2003ApJ...596..561M, 2007ApJ...658.1217M} and on-line sources\footnote{see http://web.mit.edu/ajb/www/browndwarfs/}.
The white dwarf spectra with a mass  range of M$_{\rm {WD}} = 0.6 - 1.1$\,M$_\odot$ were used with a  0.1\,M$_\odot$  step and  the  radii were   calculated using the white dwarf radius-mass relation of  $$R_{\rm {WD}} = 1.12\times 10^9\left(1-\frac{M_{\rm {WD}}}{1.44M_{\odot}}\right)^{\frac{3}{5}}$$ from \citet{1972ApJ...175..417N} and \citet{1995CAS....28.....W}.  
Spectra of WDs in  the 4000-25000\AA\  range with   pure hydrogen atmosphere were obtained using ATLAS9  \citep{1993KurCD..13.....K} and SYNTH \citep{1992stma.conf...92P} codes  for an appropriate  range of temperatures. Although our previous temperature estimate was $T_{\rm {WD}}=15\,600\pm1\,000$K based on fits to  the absorption portion of Balmer lines  \citep{2006A&A...449..645Z},  in the present fitting procedure we allowed a wider temperature range   from  $T_{\rm {eff}}=$11\,000 to 18\,000\,K,  because  of the  larger number of free parameters.  
 The calculations were performed with a 1\,000\,K  step and with the surface gravity $g= \gamma \frac{M_{\rm {WD}}}{R_{\rm {WD}}^2}$.  The spectra are  normalized to $\lambda_0 = 5500 \mathrm{\AA}$ and the contribution of the WD is 
$$ F_{\rm {WD}}(T_{\rm {eff}}, \lambda) = C_1(\delta) * F^{\rm {norm}}_{\rm {WD}}(T_{\rm {eff}}, \lambda),$$
where $C_1(\delta) = 10^{-0.4*(V+\delta+M^0_V)}$,  V = 17.8 is the object's magnitude in quiescence,  and  $\delta$ is a parameter, determining the  contribution of the  flux from the  WD in the  V band. Finally, the $M^0_V=21.109$ - is the constant to convert   magnitudes into flux (in $ergs/ cm^2/s / \mathrm{\AA}$)  in the  V band. 
The spectra of the accretion disk  was assumed to be   a simple power law
 $$F_{\rm {AD}}(\lambda) = (C_1(0)- C_1(\delta))\times\left(\frac{\lambda}{\lambda_0}\right)^{-\frac{7}{3}},$$
where $ (C_1(0)- C_1(\delta))$ determines the contribution from the accretion disk in the V band, assuming that the WD and the accretion disk are the only contributors in that wavelength  as the only other contributor is the brown dwarf,  which  has a negligible flux in V.

 The distance to the object is estimated  to be  $$ d = R_{\rm {WD}} \sqrt{\frac{F^{\rm {bb}}(T_{\rm {eff}}, 5500\mathrm{\AA})}{F_{\rm {WD}}(5500\mathrm{\AA})}}, $$
where $F^{\rm {bb}}(T_{\rm {eff}}; 5500\mathrm{\AA})$ is the black body flux at
$\lambda=5500\mathrm{\AA}$ with effective temperature $T_{\rm {eff}}$.

Observed SEDs of red/brown dwarfs of spectral types between M6 to L5 normalized to fit the observed flux in J were used. The bolometric correction to the J magnitude for each spectral type was taken from \cite{Tinney}.

\begin{figure}
  \includegraphics[width=17cm,clip=]{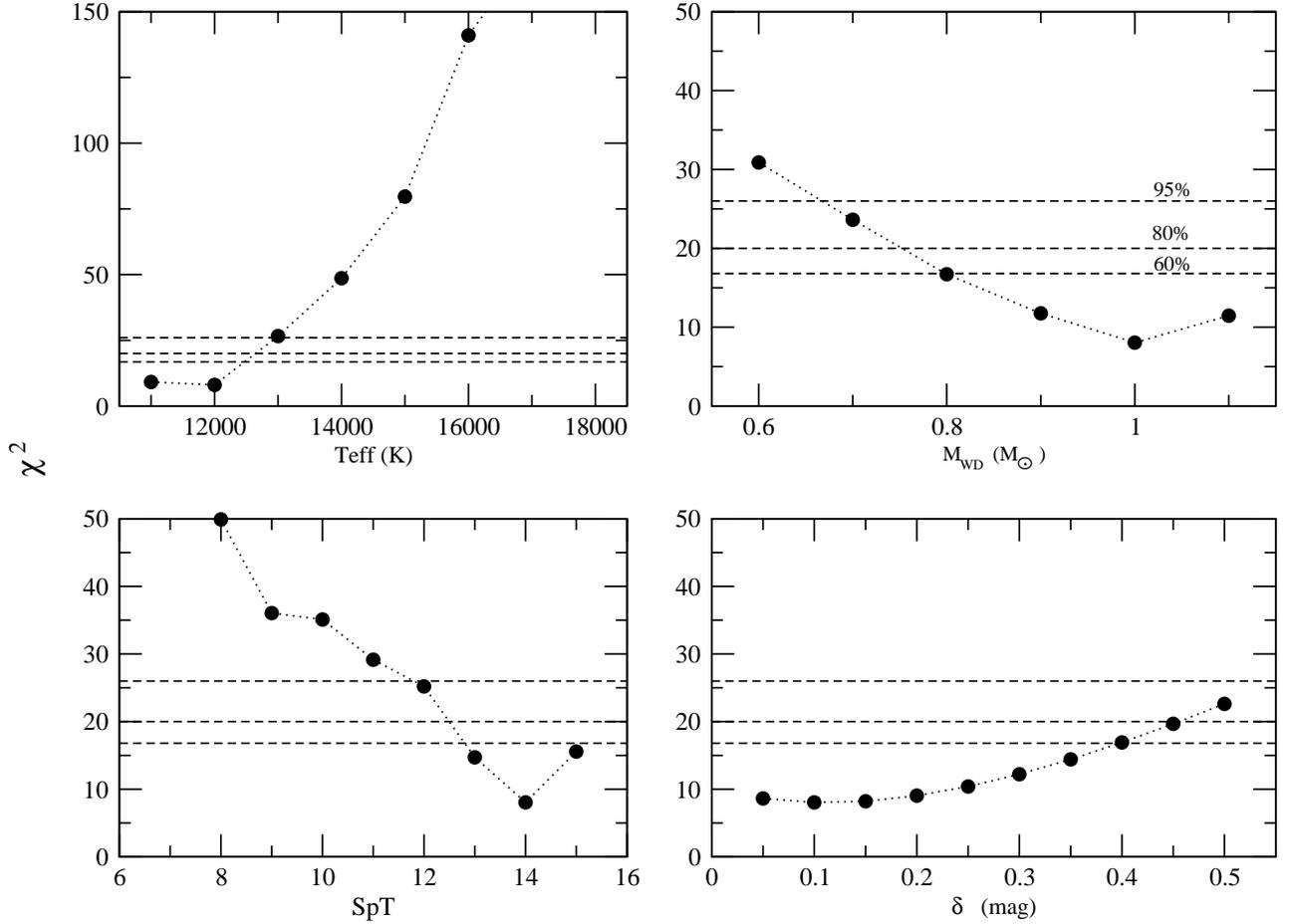}
   \caption{$\chi^2$ vs  parameters of the fit, where T$_{eff}$ is an effective temperature of the primary WD, M$_{WD}$ is  a  mass of the WD, SpT is a spectral type of the secondary (from M6V/SpT=6 to L6/SpT=16) and  $\delta$  is a ratio between accretion disk and WD contribution in the continuum flux of the SDSS1238 in the V band.  The formal confidence levels by the fit are presented by dashed lines. The numbers at the dashed lines are a probability
   to reject a model with $\chi^2$ above corresponding line.} 
   \label{fig:3}
\end{figure}

The free parameters of the three-component model are: the white dwarf effective temperature, $T_{\rm {eff}}$ , the mass of the white dwarf, $M_{\rm {WD}}$, 
the spectral type of the secondary star, SpT, and   the parameter  $\delta$ (in magnitudes).
The best fit  model to the observed  SDSS1238 spectrum
in the 0.4-2.5$\mu$m range 
 achieved for the following set of parameters (Fig.\ref{fig:1}) is: $T_{\rm {WD}}=12000K$, SpT = L4, $M_{\rm {WD}}=1.0M_{\odot}$ and $\delta=0.1$. 
 The deduced distance to the object is 110\,pc.  We studied the behavior of $\chi^2$ vs a single fit parameter, when the other  three are fixed to  their corresponding best values. 
 Fig.\ref{fig:3} presents  $\chi^2$ plots for various parameters with  marked confidence levels corresponding to 95, 80 and 60\%.   We can state that  $\chi^2$ tends to the minimum value always, when  $T_{\rm {WD}}=12\,000$\,K and $\delta = 0.05$ --- $0.15$ regardless of the value of the other parameters. At the same time, there is a dependence between the mass of the WD and the spectral type of the secondary:
the lower the mass, the larger  the radius of  WD, and thus, the  larger  the distance to the system,  resulting in an earlier spectral type of the secondary. The best fit to the optical part of the spectrum is reached with the mass of the WD of    $M_{\rm {WD}}=1.0\,M_{\odot}$, leading to the cited distance of, 110\,pc, and spectral type of the secondary, L4.  The entire range  of secondary from  $SpT=M9$ at $M_{\rm {WD}}=0.6M_\odot$ and $d = 160$pc  to $SpT=L4$ at $M_{\rm {WD}}=1.0\,M_\odot$, $d = 110$pc was considered. However,  a pronounced minimum in $\chi^2$  for rather massive WD strongly suggests  the presence of  brown dwarf in this system.  The indication of a massive WD is not accidental, a CV with a brown dwarf secondary, supposed to have evolved beyond minimum orbital period limit, has an age $\sim 3-5 \times 10^{9}$ years and a long history of accretion. 
 A priori, the bounce-back CVs are assumed to harbor a massive WD, although the distribution of  known masses of all WDs in eclipsing systems does not show any trend \citep{2006MNRAS.373..484K}. But when considering only the short-period end of that distribution as \citet{2008MNRAS.388.1582L} did, then it becomes apparent that evolved systems systematically have more massive WDs.  The upper limit for the mass of the secondary is $M_{\rm {BD}}\leq0.09M_{\odot}$ for $SpT = M9$ or $M_{\rm {BD}}\leq0.07M_{\odot}$ for $SpT = L4$ \citep{Close}. But it is also a well known fact that the secondaries of CVs tend to show systematically an earlier spectral class and a larger radius with secondaries of lower mass than the corresponding single stars on the main sequence \citep{2008MNRAS.388.1582L}.  
So we expect that the mass of the secondary in \sdss\ hovers around the lower edge of the above mentioned range of masses.  

Note, that due the high galactic latitude of this object ($b=59.5$) and the inferred small distance, the interstellar extinction is negligible.  We also would like to emphasize that in the  above calculations we took into account the fact that the state of the system was unknown at the moment of acquiring the IR data. Therefore, we conducted the fitting for both cases, considering that the system might have been at the maximum of the \brs\  during the IR observations or at the bottom. That introduced only a minor change, basically decreasing the distance to the system by $\sim15$pc and not affecting our conclusions regarding  the spectral type of the secondary.  Finally, we would like to comment on the discrepancy  in the temperature determination of the WD, which is 
 larger than we would like as compared to that of  \citet{2006A&A...449..645Z}.  In the latter  the temperature and gravity could not be determined simultaneously, and one parameter had to be fixed in order to calculate the other,  which always introduces ambiguity, as none of these two parameters could be estimated independently. In this paper we take a more complex approach:  not only profiles of the lines are being fitted, but the continuum is taken into consideration as well. Also,  the presence of the secondary, adding additional restraints on the distance and thus, on the size of the WD. The SED of the accretion disk, which may not necessarily obey the canonical power law index  is the source of the largest uncertainty in our analysis. Since the contribution of the disk is limited to 
only $\leq20$ \%, so is the accuracy of our estimates. With all that in mind, we still end up with a range of spectral classes for the secondary 
that  implies that \sdss \  is a bounce-back system.

\section{Doppler tomography  of SDSS1238}
\begin{figure}
  \includegraphics[width=18cm,bb = 130 180 545 545, clip=]{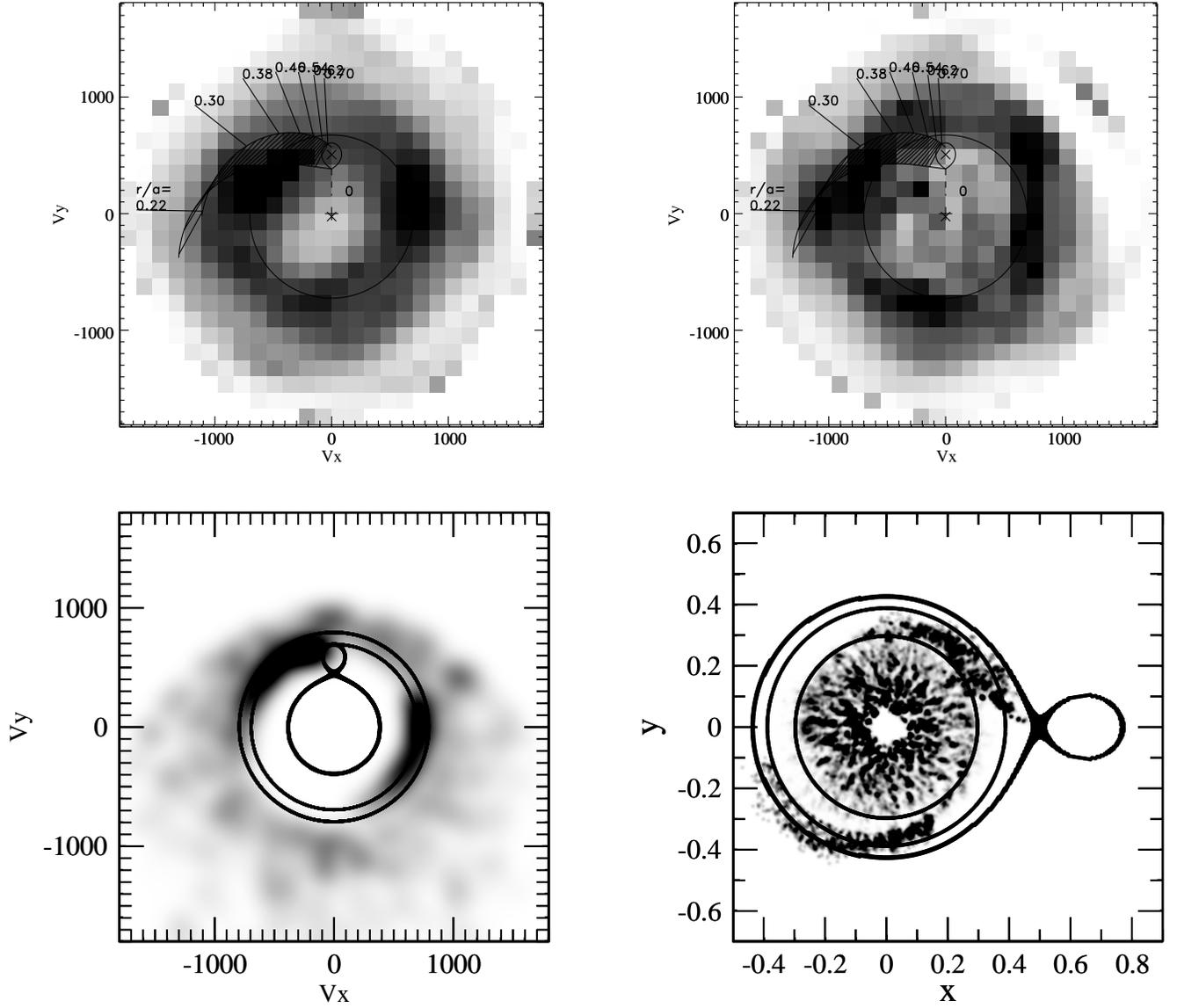} 
   \caption{top) H$_\alpha$ Doppler maps constructed on all data obtained in the 2004  (left) and 2009 (right) yy. The circle show the velocity at 2:1 resonance radius.  bottom) The synthetic  Doppler map  (left) obtained from a model accretion disk (right) for the system with M$_{WD}=1.0M_\odot$ and $q=0.05$. The circles correspond to 2:1 and 3:1 resonance radiuses.   
    } 
   \label{fig:dop}
\end{figure}

The H$_\alpha$ emission line originating in  the accretion disk  is the least affected by the absorption from the underlying white dwarf. Therefore, we constructed   Doppler maps \citep{Marsh} of  H$_\alpha$  using  all our  data obtained in  2004 and 2009  (Fig.\ref{fig:dop}, top). The individual maps of separate nights resemble each other so much, that combining all available data did not smooth out details, but made them much more convincing.  In order to over-plot  contours of the secondary star, the location of the WD, the trajectory of the stream and resonance radius of the disk, we used the best fit parameters to the SED (see previous section). Thus, the white dwarf mass of  $M_{\rm {WD}} = 1.0\,M_\odot$  was adopted. Due to the ambiguity in determining    precise masses of brown dwarfs,   we adopted  a mass ratio of $q = 0.05$. This value is   typical  for systems considered as bounce-back \citep{2006MNRAS.373..484K} and it is the same mass ratio which was obtained for  SDSS0804, a twin of \sdss,  from the super-hump period observed during  the super-outburst in 2006 \citep{2007ASPC..372..511P,2008A&A...486..505Z}. Before  SDSS\,0804  underwent a WZ-Sge type super-outburst,  it  showed   similar   peculiar photometric  variability  to \sdss. 
In addition, it shares   every other characteristic of a  WZ Sge-type object, and was also proposed as a candidate to the bounce back system \citep{2008A&A...486..505Z}.

As already noted,  the structure of the accretion disk did not change between the two epochs of our observations. There is a bright  spot  at the expected place where the stream of matter from the secondary collides with the accretion disk, but it overlaps with a much larger and prolonged structure, too extended to be a part of the spot.  Another  extended bright region of similar size can be seen  at velocity coordinates  ($\approx700$ km s$^{-1}$, $\approx 0$~km~s$^{-1}$) as well as  a less bright structure at ($\approx -200$ km s$^{-1}$, $\approx -800$ km s$^{-1}$).  Similar Doppler maps were obtained for WZ Sge during super-outburst in 2001 \citep{2002PASJ...54L...7B, 2003A&A...399..219H, 2004RMxAC..20..178S}
and in quiescence for  bounce back candidates SDSS\,1035 \citep{2006MNRAS.373..687S} and  SDSS\,0804 \citep{Andres}. 
 
  Such brightness distribution in the Doppler map can be interpreted as  evidence of spiral waves in the disk (see for example, \citet{1999MNRAS.307...99S} and  \citet{2001LNP...573...45S} and reference therein).   The formation of a spiral structure in an accretion disk of  a close binary system was predicted by \citet{1979MNRAS.186..799L} and explored by various authors  (\citet{1990A&A...235..211M, 1994A&A...288..807H, 1999MNRAS.304..687S, 2005ASPC..330..389K, 2007MNRAS.376...89T} and reference therein). 
\citet{1986MNRAS.221..679S, 1986MNRAS.219...75S} demonstrated from high resolution numerical calculations that spirals will always formed in  accretion disks under tidal forces from the secondary. They actually used $q=1$ in their models, but observationally, such spirals were detected  in a number of systems only  during outbursts of dwarf novae. The careful examination of quiescent disks of the same systems did not reveal any spiral structures in longer period DNe. \citet{1999MNRAS.307...99S} argued that little evidence of spiral arms in the emission lines is expected in  systems with low  values of viscosity.

On the other hand,  spiral arms related to  2:1 resonance can be found in  systems  with extremely low mass ratio $q< 0.1$  as originally was predicted by \citet{1979MNRAS.186..799L}. The bounce-back systems and related to them, WZ\,Sge  stars, are examples of such objects. The long outburst recurrence time in WZ\,Sge systems is probably explained by a very low viscosity in their accretion disks, yet 
 spiral arms can be observed permanently in quiescent bounce-back systems in which, on one side there is a massive WD, which gained mass 
 during a long accretion history, and on another side there is a late-type brown dwarf, providing a mass ratio of  $\le 0.06$. 
Fig.\ref{fig:dop}  (bottom, left) depicts a synthetic  Doppler map  constructed from a model accretion disk that is shown on the bottom right panel. The latter was calculated with  the binary system parameters derived above by using smooth particle hydrodynamics according to 
\citet{1996MNRAS.279..402M, 1997MNRAS.289..889K}. The artificial doppler map reproduces the observed map in a case when there is a brightness excess within  spiral arms.  Most of the disk particles are on periodic orbits, which are most favorable from the point-of-view of viscosity. However, the resonance dispatches some particles onto aperiodic orbits creating viscosity perturbations, which will create excess of heat. A slightly different interpretation of spiral arm brightness is offered  by \citet{2002MNRAS.330..937O}.  The mechanism is not very well established, but it is natural to assume that in these regions there will be excess emission. 

\section{Cyclic brightenings.}
\begin{figure}
  \includegraphics[angle=0,width=17cm, clip=]{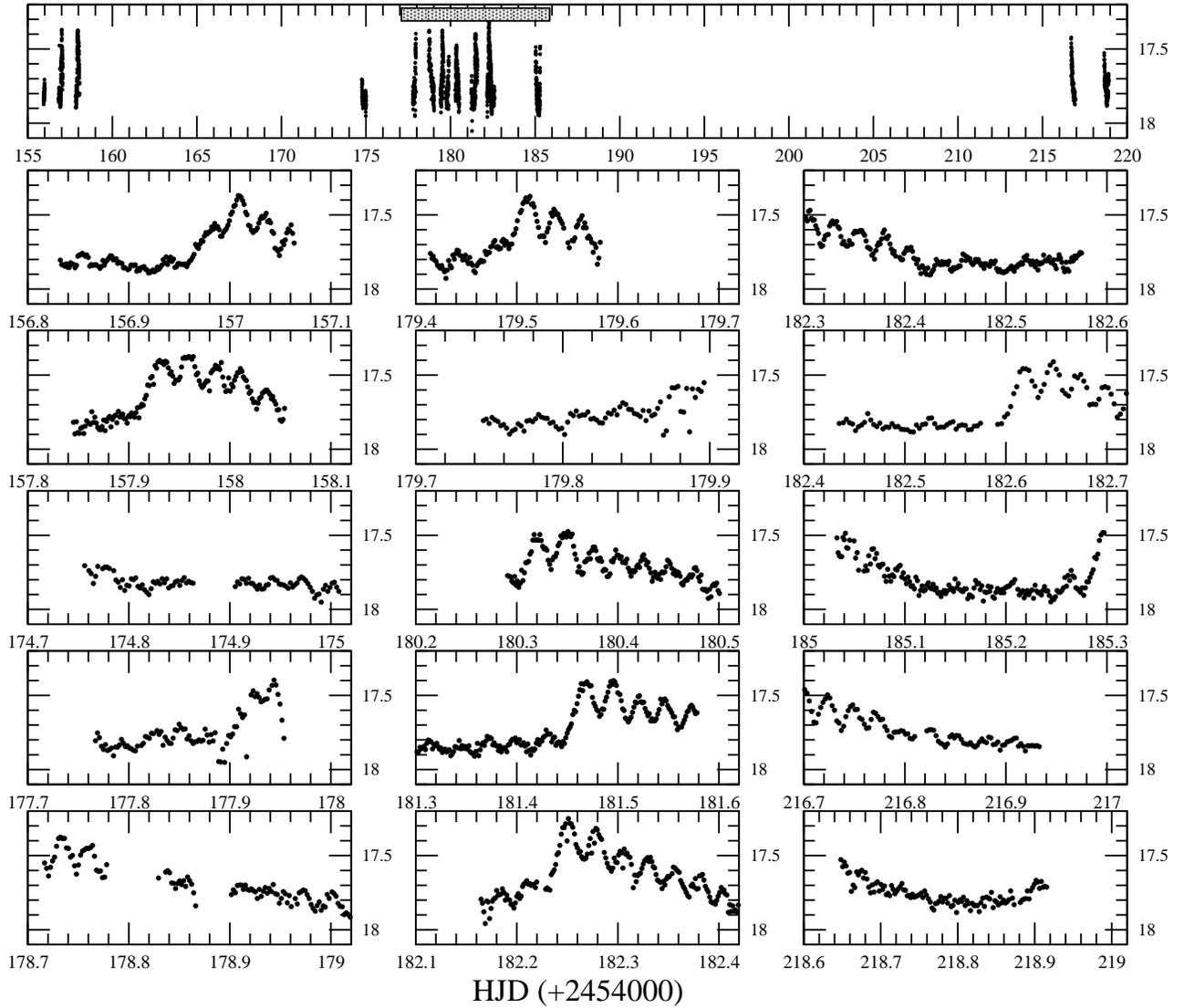}
  \caption{The light curve of SDSS 1238 in V band obtained in the 2007 year. Each night is presented in a separate panel.  
  }
  \label{fig:phot}
\end{figure}

\begin{figure}
  \includegraphics[angle=0,width=17cm, clip=]{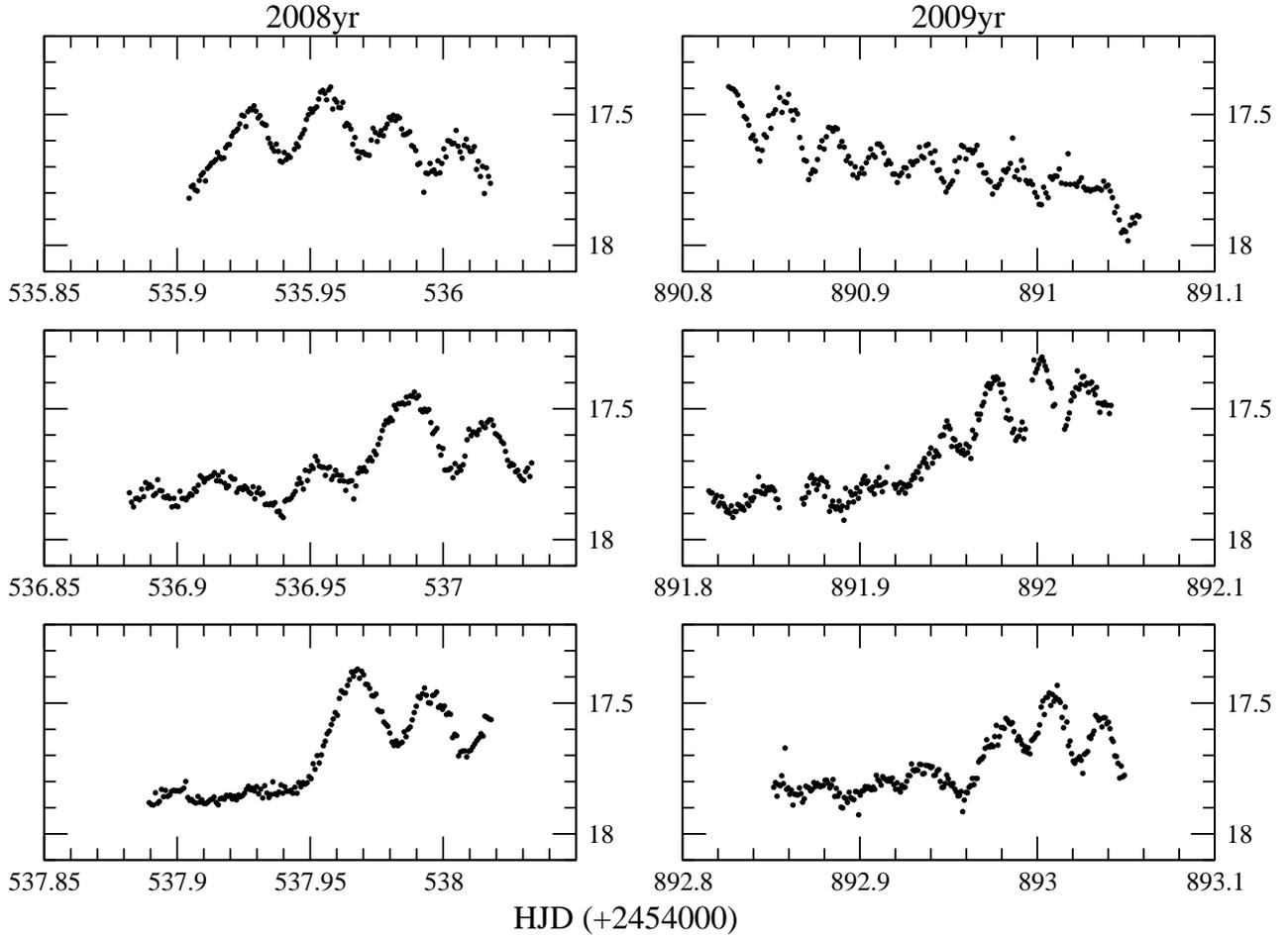}
  \caption{The light curve of SDSS 1238 in V band obtained in  2008 and 2009 yy. Each night is presented in a separate panel.}
  \label{fig:photNew}
\end{figure}

\begin{figure}
  \includegraphics[angle=0,width=17cm, clip=]{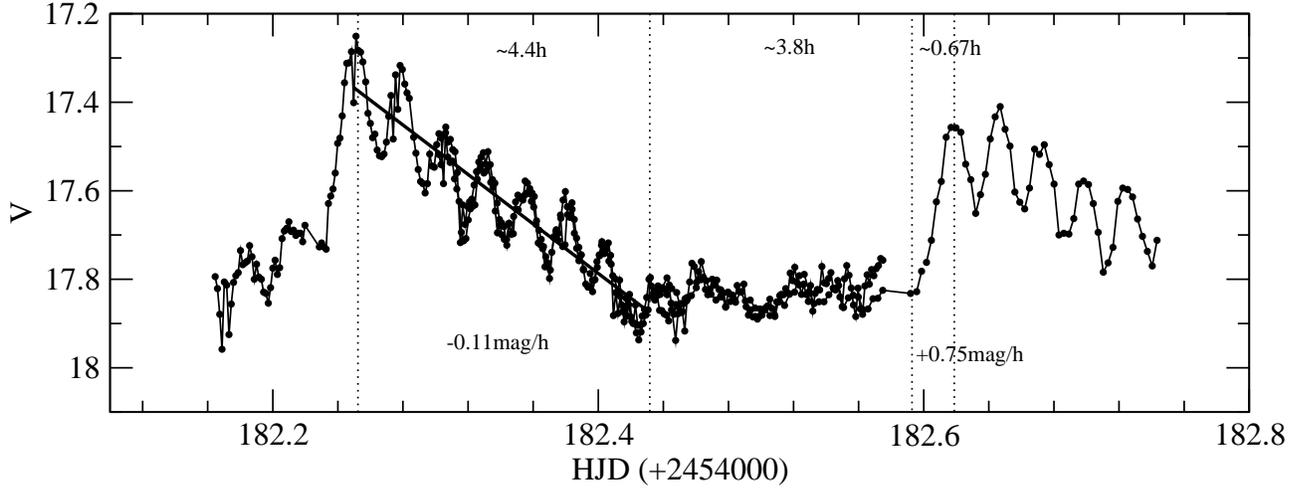}
  \caption{The light curve of SDSS 1238 in V band obtained during about 15 hours continuos observations  in HJD 2454182. Vertical dashed lines select different phases of the long-term variability (LTV). The numbers in the top of the figure are approximately  durations for each marked phase of LTV and the numbers in the bottom are corresponding rates of the magnitude change during fading and increasing of the object brightness. }
  \label{fig:phoCont}
\end{figure}

\begin{figure}
  \includegraphics[width=17cm, clip=]{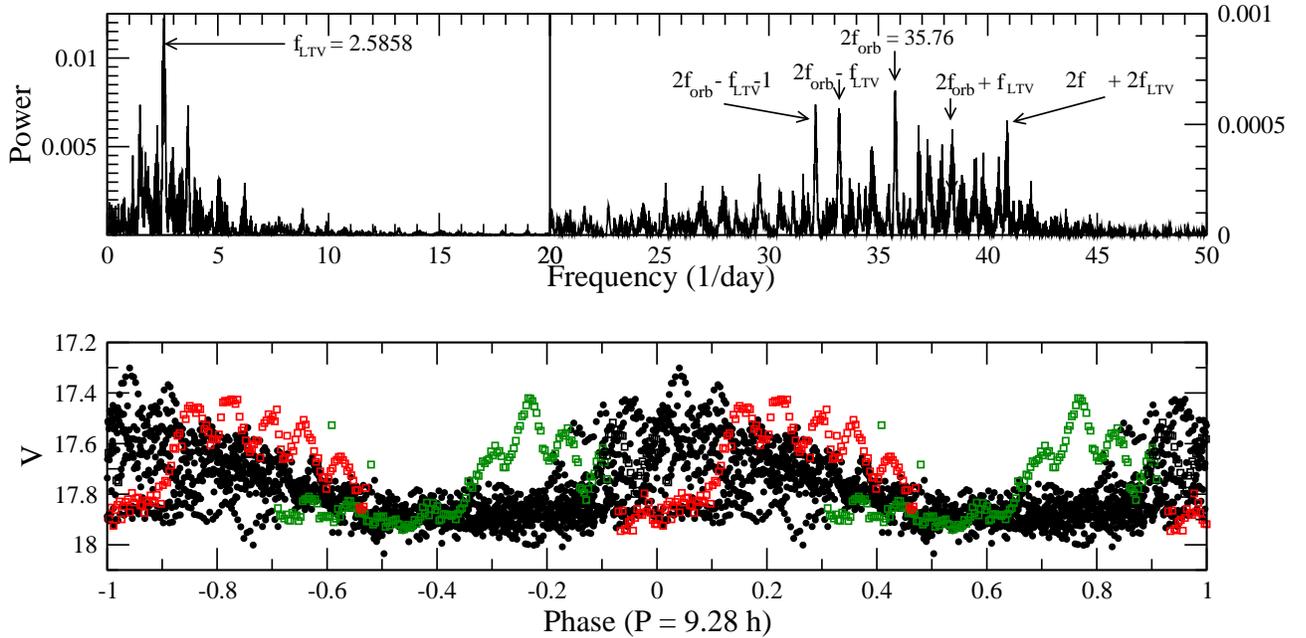}
  \caption{top) The power spectrum of the all photometric data presented in the top panel of Fig.\ref{fig:photNew}.
  bottom) The light curve comprised of the all data folded on $P=9.28$h.  The open color points show {\it brightenings} with maximal displacment in the light curve folded on $P=9.28$h.  }
  \label{fig:LTV1}
  \end{figure}

\begin{figure}
  \includegraphics[width=17cm, clip=]{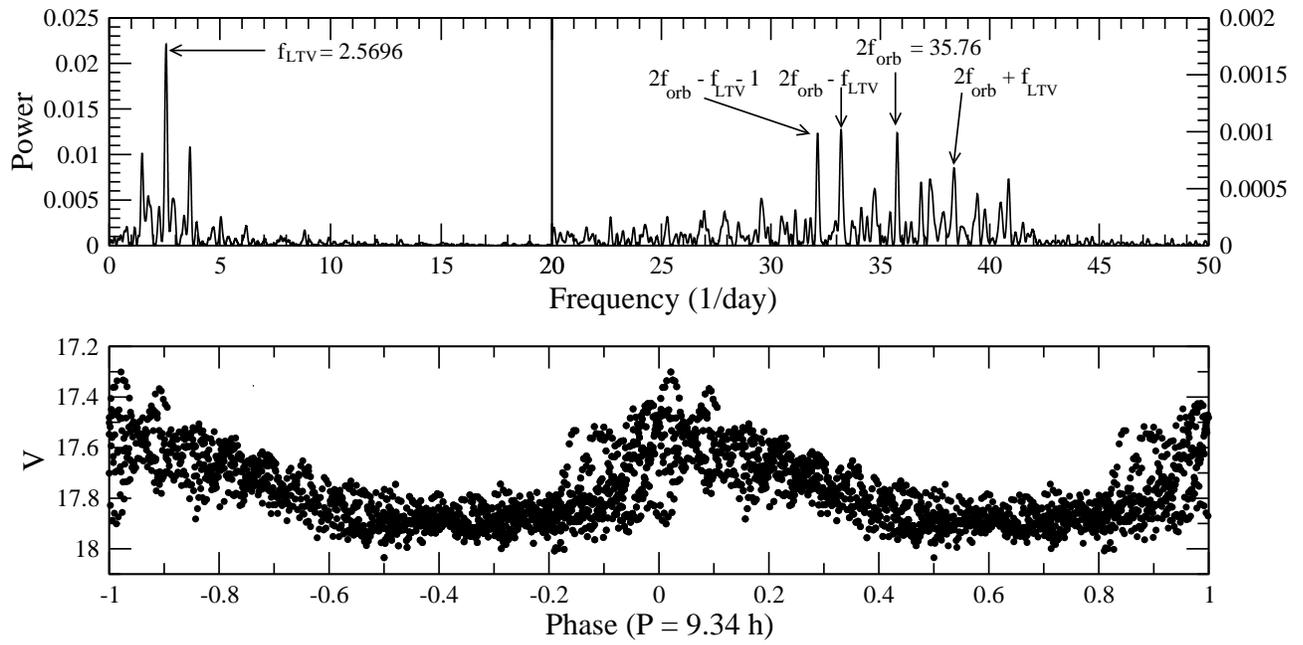}
  \caption{top)  The power spectrum of the data obtained  in the period of HJD 2454177.1-2454185.6 (see Fig.\ref{fig:photNew}).
  bottom)  The light curve comprised of the selected data from the period of HJD 2454177.1-2454185.6 folded on $P=9.34$h period.}
  \label{fig:LTV2}
  \end{figure}

Figure \ref{fig:phot} displays the light curve of SDSS\,1238 obtained  in 2007. In general,  the object shows an identical behavior to previous years,  as described by \citet{2006A&A...449..645Z}.   There are two distinct types of variability: a long-term variability (LTV), lasting more than 8h, and a short-term variability with a period corresponding to half  the orbital period. Follow-up observations during 2008 and 2009  confirm a steady presence of both types of variability in the light curve (Fig.\ref{fig:photNew}).
Continuous observations during about 15 hours obtained on 22/03/2007  (HJD 2454182.15-2544182.60, Fig.\ref{fig:phoCont}) allowed us to observe two consecutive {\it brightenings} and, thus  to determine their recurrence time  directly ($\sim 9$h).   \citet{2006A&A...449..645Z} demonstrated that \brs\ occurred cyclically with periods in between 8 and 12 hours, however it was not possible to establish if the phenomenon was strictly periodic or not. Armed with more data, and the advantage of detecting consecutive events by employing multi-longitude observations, we performed a simple period search. The period analysis  of all  data obtained during 2007, based on a discrete Fourier transform method \citep{1975Ap&SS..36..137D},  results in  a strong peak at a frequency  $f_{\rm {LTV}}=2.59$ day$^{-1}$ with FWHM of 0.1. This  conforms with the period of $P_{\rm {LTV}}=9.28\pm$0.36h $\approx 7 P_{\rm {orb}}$ (Fig.\ref{fig:LTV1}). However,  the data  folded with this period looks messy  because, apparently, some re-brightenings  happen with a different cycle period.   It shows that the large uncertainty in the period value is not just a result of a scarce amount of data, or its uneven  distribution, but that the \brs\ are not  strictly periodic in long timescales.   The peak $2\times f_{\rm {orb}}=35.76$\ day$^{-1}$, corresponding to the half of  the orbital period, $P_{\rm {orb}}/2=40.3$ min, is also present in the power spectrum. There are some additional  peaks which are one-day aliases or  a combination of high harmonics of $f_{\rm {LTV}}$ with $2\times f_{\rm {orb}}$.  We have repeated the same analysis for the data obtained only within HJD 2454177.1-2544185.6, when we observed the object with small observational gaps. The result of the analysis of  periodicity for this selection is presented in Fig.\ref{fig:LTV2}. The strongest peak in the power spectrum appears at a frequency corresponding to  $P^{*}_{\rm {LTV}}=9.34(26)$h. The light curve folded with this period  is  coherent during the considered time (Fig.\ref{fig:LTV2}),  i.e macro-profile of the \brs\  (the rise, the peak, the declining slope and minimum duration) repeats itself with high accuracy and, if not for the high-frequency modulation also present  in the light curve, the folded curves would concur.  Combining the entire available data obtained during the 2004 --- 2009 observational runs   does not permit a determination  of  a single period with which the entire data set can be folded.  Clearly,  the LTV modulation is not strictly periodic, but probably  maintains coherence  in  a time scale of  several weeks and certainly has a cyclic nature. 
  
 The  long-term variability ({\it the brightenings}) presents a fast rise in the object brightness of up to 0.4-0.6 mag, lasting about half  the orbital period  ($\sim 0.7$\,h) with the brightness falling back to the quiescence level  during the next 4.4-4.6\,h (Fig. \ref{fig:phoCont}). After that, the object  remains in quiescence during about another 4 hours.  The brightness increases at a   rate  of  $\sim0.75$ mag/h, and  the falling  rate is about $\sim0.10$ mag/h. Sometimes, the rise in the brightness happens more slowly and  lasts until almost one full orbital period (Fig.\ref{fig:phot} nights: 156, 179).  

A short-term variability  is clearly present during {\it the brightenings}.  The amplitude of 
the short-term variability depends strongly on the phase  of  {\it  brightenings}. The amplitude is larger then  the total brightness increases and falls, sometimes  practically disappearing, between  the \brs\  (Figs. \ref{fig:phot}, \ref{fig:photNew}, \ref{fig:phoCont}). 
Curiously, when the LTV is coherent, the signal at double orbital frequency decreases and a beat frequency between the LTV and $2\times f_{\rm {orb}}$ becomes  dominant in the range of frequencies around 36 cycles/day. Removing LTV has little effect, because this does affect the   frequency modulation but not the  amplitude modulation. The strong relation between LTV cycle coherence and the strength of $2\times f_{\rm {orb}}$ modulation  attests that these two phenomena are physically linked in more than a simple amplitude correlation. We suspect that the \brs\  and formation of spiral arms in the disk are the  result of the same process, but its   cyclic nature remains  unclear.

\section{Discussion and conclusions}
\label{sec: dics}

In recent years, a number of objects were discovered, mostly thanks to the Sloan Digital Sky Survey, which can be characterized  as WZ Sge type CVs in quiescence,  judging by their $\sim 80-85$ min periods, spectral appearance and lack of SU UMa style outburst activity.  Currently, there are about 20 objects with similar characteristics which were selected from the SDSS lists. Three  such objects  are   GW Lib, 
V455 And and SDSS0804,  which underwent a typical WZ\,Sge type super-outburst in 2006-2007, confirming the correct assessment of their nature from their behavior in quiescence.  But unlike classical WZ Sge stars, some of them differ by the presence of permanent double-humps per orbital period in the  light curve. The significance of the double-humped light curve lies in the assumption that it is produced by the spiral structure in the accretion disk.  The spiral arms in accretion disks have been detected before  by means of Doppler tomography in   long period systems only during  DN outburst. We are not aware of any reports of double-humped light curves during these detections.  Double-humped light curves have previously came to attention, as they were detected during super-outbursts of WZ Sge. It was soon suggested that they are result of 2:1 resonance \citep{2002A&A...383..574O, 2002PASP..114..721P}, as one of the possibilities. Also spirals arms are formed in the accretion disk of the WZ Sge   undergoing super-outburst \citep{2004RMxAC..20..178S}, because, according to our hypothesis WZ Sge accretion disk reaches the resonance radius only during super-outburst.  Nevertheless, the  spiral structure has been observed persistently  along with the double-humped light curve  in two similar short-period  systems, SDSS\,0804 and SDSS\,1238 in quiescence. We assume that the spiral structure formed in low viscosity, quiescent disks is a result of  distortion of the disk by the 2:1 resonance. The 2:1 resonance  may happen only in the  accretion disk of a  system with extreme mass ratio of $q \le 0.1$. Such mass ratio is achieved only in systems known as bounce-backed or, in other words, systems which have reached  a minimum period of $\approx 80$ min and have turned to  slightly longer periods according to \cite{1981AcA....31....1P}.    
 It is expected that bounce-back systems are numerous \citep{1999MNRAS.309.1034K}, but until recently very  few candidates have been found. The SDSS helped to uncover a large number of new CVs, with new interesting features \citep{2009MNRAS.397.2170G}. Among them, there is  a number of short period systems, some of which turned  out to belong to the long-sought bounce-back systems \citep {2008MNRAS.388.1582L, 2002MNRAS.336..767M}. 

We have unveiled a  brown dwarf secondary in \sdss, probably as late as L4, thus providing strong evidence that this object is a real bounce-back system. We also estimated parameters of the WD. The best fit to the SED converges if the primary  is a massive   $M_{\rm {WD}}\approx1.0\,M_{\odot}$ white dwarf with a $T_{\rm {WD}}=12\,000\,$K temperature. Both numbers seem plausible since  the system is very old and the WD is expected to be relatively massive and cool. These findings provide further support to our claim that   by simply observing a permanent double humped light curves one can identify an  evolved, bounce-back systems instead of using other complicated methods.

It is important to note that  \sdss\ has another peculiarity, which \citet{2006A&A...449..645Z} termed as \brs. There was only one other system known to exhibit \brs, namely SDSS\,0804, but since it underwent a super-outburst in 2006, its photometric behavior has drastically changed. 
The \brs\ shortly detected by \citet{2006AJ....131..973S} in SDSS\,0804 before the super-outburst have been replaced by:  a) {\it the  mini-outburst}  activity with permanent presence of the  double-humped light curve of constant amplitude \citep{2008A&A...486..505Z} and  by: b) a 12.6 min period,  probably corresponding to pulsation activity of the WD \citep{2007ASPC..372..511P}.
Therefore, at present, \sdss\ is the only object known to show \brs.  It greatly complicates the study of this phenomenon. Based on new multi-longitude continuous monitoring, we demonstrated here that the \brs\ are of cyclic nature with a recurrence time of $\approx9$ hours and they are probably coherent over several cycles. There seem to be  strong correlation between  \brs\  cycles and the  amplitude of double-hump periodic variability. This   is believed to be the result  of spiral arms in the accretion disk of bounce-back systems with an extreme mass ratio, in which accretion disk extends beyond a 2:1 resonance radius. The tomogram of a simulated accretion disk in the regime of resonance, closely resembles the observed one and supports this hypothesis.  The formation of a spiral structure in the disk can be accounted by the appearance of double humps in the light curve, but it can not be the reason of increased brightness of the disk. The disk brightness  directly depends on the mass transfer rate and its change should probably reflect change in mass transfer.  No readily explanation is available as to why that rate can be variable and cyclical, but possible speculations include counteraction: a) to the heating of the secondary by \brs, or b) to tidal interaction between  the secondary with the resonance attaining accretion disk.

\acknowledgments

This work was supported in part be DGAPA/PAPIIT projects IN109209 and IN102607.

\clearpage


\clearpage
\begin{table}
\caption{Log of time-resolved observations of SDSS J123813.73-033933.0 in V band}
\begin{tabular}{llllccc}
\tableline\tableline
 Date        & HJD Start+ & Telescope & Exp.Time   & Duration\\
 & 2454000 &   &                 Num. of Integrations   &                          &                \\  \hline
 Photometry &  &   &    &                   &                          &                \\ 
 25 Feb. 2007 & 156.832 &1.5m/SPM/M\'exico       &120s$\times$160                    & 5.3h  \\
26 Feb. 2007 &  157.845  &1.5m/SPM/M\'exico     &120s$\times$144                  & 4.8h  \\
15 Mar. 2007 & 174.756  &1.5m/SPM/M\'exico      &220s$\times$95                   & 5.8h  \\ 
18 Mar. 2007 & 177.767  &1.5m/SPM/M\'exico      &170s$\times$91                   &4.3h  \\
19 Mar. 2007 & 178.716  &1.5m/SPM/M\'exico      &240s$\times$108                   &7.2h  \\ 
20 Mar. 2007 &  179.414 &2m/Terksol/Russia        &120s$\times$118                   &3.9h  \\ 
20 Mar. 2007 & 179.746  &1.5m/SPM/M\'exico      &160s$\times$77                    &3.4h  \\ 
20 Mar. 2007 & 180.290 &2m/Terksol/Russia         &120s$\times$152                    &5.1h  \\ 
21 Mar. 2007  & 181.169  &2m/IAO/India                &120s$\times$135                    &7.5h  \\ 
21 Mar. 2007 & 181.286 &2m/Terksol/Russia         &120s$\times$210                    &7h  \\ 
22 Mar. 2007  & 182.164  &2m/IAO/India                &120s$\times$198                    &6.9h  \\ 
22 Mar. 2007 &  182.302 &2m/Terksol/Russia        &120s$\times$195                    &6.5h  \\ 
22 Mar. 2007 &  182.435  &0.8m.IAC80/Spain      &270s$\times$98                     &7.4h  \\ 
25 Mar. 2007 &  185.032  &2.1m/BOAO/Korea       &135s$\times$107                    &4.0h  \\
25 Apr. 2007  &  216.696  & 0.8m/IAC80/Spain    &200s$\times$101                    &5.6h  \\
27 Apr. 2007 &  218.648  & 0.8m/IAC80/Spain     &200s$\times$116                    &6.4h  \\ \tableline
09 Mar. 2008 & 535.904 & 1.5m/SPM/M\'exico      &60s$\times$150		            &2.5h\\
10 Mar. 2008 & 536.882 & 1.5m/SPM/M\'exico      &60s$\times$200		            &3.3h\\
11 Mar. 2008 & 537.889 & 1.5m/SPM/M\'exico        &60s$\times$170		            &2.8h\\ \tableline
27 Feb. 2009 & 890.826 & 1.5m/SPM/M\'exico        &60s$\times$184		            &3.1h\\
28 Feb. 2009 & 891.814 & 1.5m/SPM/M\'exico        &60s$\times$227		            &3.8h\\
01 Mar. 2009 & 892.851 & 1.5m/SPM/M\'exico        &60s$\times$218		            &3.6h\\ 
\tableline
\end{tabular}
\end{table}

\clearpage
\begin{table}
\caption{Log of time-resolved spectroscopic   observations of SDSS J123813.73-033933.0}
\begin{tabular}{llllccc}
\tableline\tableline
Date        & HJD Start+ & Telescope & Exp.Time& Duration\\
 & 2454000 &   &    &                 Num. of Integrations   &                          &                \\  \tableline

 Spectroscopy &             &                                   &                                                 &                              \\ 
 24 Jan. 2009  & 855.99 &  2.1m/SPM/M\'exico & 486s$\times$7       &    0.52h       \\
 25 Jan. 2009  & 856.96 &  2.1m/SPM/M\'exico  & 486s$\times$17     &    2.43h       \\ 
 26 Jan. 2009  & 857.90 &  2.1m/SPM/M\'exico & 486s$\times$20         &   2.69h      \\
 27 Jan. 2009  & 856.96 &  2.1m/SPM/M\'exico  & 486s$\times$10       &    1.2h     \\ \tableline

\end{tabular}
\end{table}

\clearpage 



\label{sec:data}

\end{document}